\documentclass[12pt,prl,aps,amssymb,amsmath,tightenlines]{revtex4} 

\usepackage[dvips]{graphicx}
\usepackage{dsfont}

\newcommand{\re}{{\rm e}}
\newcommand{\ri}{{\rm i}}

\usepackage{xcolor}

\usepackage{ulem}
\begin{document}
\title{Comment on `Comment on ``Hamiltonian for the zeros of the Riemann zeta
function''\,'}
\author{Carl M. Bender$^1$, Dorje C. Brody$^{2,3}$, Markus P. M\"uller$^{4,5}$}

\affiliation{
$^1$Department of Physics, Washington University, St Louis, MO 63130, USA\\
$^2$Department of Mathematics, Brunel University London, Uxbridge UB8 3PH,
United Kingdom\\
$^3$Department of Optical Physics and Modern Natural Science, St Petersburg
National Research University of Information Technologies, Mechanics and Optics, 
St Petersburg 197101, Russia \\ 
$^4$Departments of Applied Mathematics and Philosophy, University of Western
Ontario, Middlesex College, London, ON N6A 5B7, Canada\\
$^5$The Perimeter Institute for Theoretical Physics, Waterloo, ON N2L 2Y5,
Canada}

\begin{abstract}
This paper is in response to a recent comment by Bellissard [arXiv:1704.02644] 
on the paper [Phys. Rev. Lett. \textbf{118}, 130201 (2017)]. It is explained that 
the issues raised in the comment have already been discussed in the paper and 
do not affect the conclusions of the paper. 
\end{abstract}

\maketitle

In a recent comment~\cite{r1}, Bellissard points out (i) that the momentum 
operator ${\hat p}$ defined on the positive half line admits no selfadjoint 
extension, and (ii) that the Hurwitz zeta function $\zeta(z,x+1)$ for
${\rm Re}(z)=\frac{1}{2}$ is not square-integrable on the positive half line.
Based on these observations, Bellissard concludes that ``While the idea is
appealing, a closer look at the paper raises several problems that are not
addressed and that make this approach quite questionable'' \cite{r1}. In what
follows we list a number of issues in \cite{r1}, showing (a) that Bellissard's 
observations are already mentioned in~\cite{r2}, and (b) that Bellissard's 
criticism does not invalidate the arguments of~\cite{r2}.

\noindent{\bf Issue 1}: The main findings of \cite{r2} are misrepresented. 
In relation to the operator 
\begin{eqnarray}
{\hat H}=\frac{{\mathds 1}}{{\mathds 1}-\re^{-{\rm i}{\hat p}}}\left({\hat x}
{\hat p}+{\hat p}{\hat x}\right)({\mathds 1}-\re^{-{\rm i}{\hat p}}) 
\label{e1}
\end{eqnarray}
introduced in \cite{r2}, Bellissard claims that ``extra discrete symmetries of
${\hat H}$ are used to show that these eigenvalues are real,'' and that ``the
authors show that it admits symmetries that force its eigenvalues to be real
leading to a potential proof of the Riemann Hypothesis.'' In fact, in \cite{r2}
it is stated that ``We are not able to prove that the eigenvalues of ${\hat H}$
are real.'' Regarding the discrete symmetry (combined parity and time reversal)
noted in \cite{r1}, it is stated in \cite{r2} that the symmetry property implies
that the eigenvalues of $\ri{\hat H}$ form complex-conjugate pairs. However,
{\it nowhere in {\rm \cite{r2}} is it claimed to have shown that the eigenvalues of
${\hat H}$ are real.} 

\noindent{\bf Issue 2}: In \cite{r1} Bellissard discusses in some detail, using the
von Neumann deficiency index argument, that the momentum operator ${\hat p}$
defined on the positive half line is not selfadjoint. This well known fact can
be found in standard textbooks (see, e.g., \cite{r3}, \S49); see also \cite{r4} 
for an elementary account. In \cite{r2}, the
nonselfadjointness of ${\hat p}$ led to a hypothesis that the action of
${\hat p}$ and that of its adjoint on the (yet to be identified) domain of
${\hat H}$ agree. Based on this hypothesis, an inner product is introduced 
that renders $\hat H$ symmetric. These formal arguments also suggest that the symmetry of $\hat p$ in this inner product follows from the boundary condition $\psi(0)=0$. Thus, the selfadjointness of ${\hat p}$ 
is never postulated or used. Indeed, the nonselfadjointness of ${\hat p}$ is already implicit in
\cite{r2}. 

\noindent{\bf Issue 3}: In \cite{r1} Bellissard uses a particular integral
representation of the Hurwitz zeta function to show that $\zeta(z,x+1)$ for
${\rm Re}(z)=\frac{1}{2}$ is not square-integrable on the positive half line. 
In fact, the asymptotic behaviour of $\zeta(z,x+1)$ is worked out explicitly in
\cite{r2}. Using this, it is shown that for ${\rm Re}(z)=\frac{1}{2}$ the
function $\zeta(z,x+1)$ behaves in magnitude like $\sqrt{x}$ for large $x$, from
which it is obvious that $\zeta(z,x+1)$ is not square-integrable. 
(An asymptotic analysis of the integral representation (2) of 
\cite{r1} leads at once to the conclusion that $\zeta(z,x+1)\approx 
x^{1-z}$ as $x\to\infty$; see also (25.11.43) in the NIST Digital 
Library of Mathematical Functions, http://dlmf.nist.gov/). 

\noindent{\bf Issue 4}: In \cite{r1} Bellissard focuses on the space
$L^2(0,+\infty)$ of square-integrable functions, which is a standard choice of Hilbert space for some situations in quantum mechanics. However, for pseudo-Hermitian operators one
must use an alternative inner product, and often the analysis requires the
identification of the appropriate inner product. The insertion of a 
weight function suggested in \cite{r1} is not a way forward. In short, whether
or not $\zeta(z,x+1)$ is an element of $L^2(0,+\infty)$ with respect to the
Lebesgue measure is not a relevant question to ask in the context of
pseudo-Hermitian operators. This fact is explained at length in \cite{r2}.

\noindent{\bf Conclusion}: In summary, the remarks of Bellissard, namely,
that ${\hat p}$ is not selfadjoint and that $\zeta(z,x+1)$ is not square-integrable, can already be found in \cite{r2}. 
It is not suggested in~\cite{r2} 
that the Hilbert space $L^2(0,+\infty)$ and its inner product can be used to make the formal arguments rigorous. Instead, the arguments in \cite{r2} rely on the techniques of 
biorthogonal quantum theory \cite{x3} appropriate for pseudo-Hermitian operators 
(`pseudo-Hermitian' in the sense of \cite{x2}). Thus, the remarks of~\cite{r1} do not invalidate the arguments of~\cite{r2}.

\vspace{0.2cm} 

DCB thanks the Russian Science Foundation for support (project 16-11-10218). MPM
is supported in part by the Canada Research Chairs program. Research at
Perimeter Institute is supported by the Government of Canada through Innovation,
Science and Economic Development Canada and by the Province of Ontario through
the Ministry of Research, Innovation and Science.


\begin{thebibliography}{999}

\bibitem{r1} J.~V.~Bellissard, Comment on ``Hamiltonian for the zeros of the
Riemann zeta function'' [arXiv:1704.02644].

\bibitem{r2} C.~M.~Bender, D.~C.~Brody, and M.~P.~M\"uller,  
Hamiltonian for the zeros of the Riemann zeta function. 
{\em Phys.~Rev.~Lett.}~{\bf 118}, 130201 (2017). 

\bibitem{r3} N.~I.~Akhiezer and I.~M.~Glazman, {\em Theory of Linear Operators
in Hilbert Space} (New York: Ungar, 1961). 

\bibitem{r4} G.~Bonneau, J.~Faraut, and G.~Valent, Self-adjoint extensions of
operators and the teaching of quantum mechanics. {\em Am. J. Phys.} \textbf{69},
322 (2001). 

\bibitem{x3} D.~C.~Brody, 
Biorthogonal quantum mechanics. 
{\em J. Phys. A: Math. Theor.}~{\bf 47}, 035305 (2014).

\bibitem{x2} G.~W.~Mackey, 
{\em Commutative Banach Algebras}. 
(Instituto de Matematica pura e Aplicada do Conselho Nacional de Pesquisa, 
Rio De Janeiro, 1959).


\end{thebibliography}
\end{document}